%% file: couplingpf.tex
\documentclass[english]{article}
\usepackage{lmodern}

\usepackage[T1]{fontenc}
\usepackage[latin9]{inputenc}
\usepackage[a4paper]{geometry}
\usepackage{dsfont}
\usepackage{color}
\usepackage{babel}
\usepackage{amsmath}
\usepackage{amsthm}
\usepackage{amssymb}
\usepackage{stmaryrd}
\usepackage{graphicx}
\usepackage{setspace}
\usepackage{esint}
\usepackage{caption}
\usepackage{natbib}
\usepackage[unicode=true,bookmarks=true,bookmarksnumbered=false,bookmarksopen=false,breaklinks=false,pdfborder={0 0 0},pdfborderstyle={},backref=page,colorlinks=true]{hyperref}
\hypersetup{pdftitle={Unbiased MCMC},pdfauthor={Jacob OLeary Atchade},linkcolor=RoyalBlue,citecolor=RoyalBlue}

\usepackage{float}
\usepackage{subfig}
\usepackage[dvipsnames,svgnames,x11names,hyperref]{xcolor}
\usepackage{nicefrac}
\usepackage{algorithm}

\floatstyle{ruled}
\newfloat{algorithm}{tbp}{loa}
\providecommand{\algorithmname}{Algorithm}
\floatname{algorithm}{\protect\algorithmname}
\newtheorem{thm}{Theorem}[section]
\newtheorem{lemma}{Lemma}[section]

\newtheorem{assumption}{\protect\assumptionname}
\providecommand{\assumptionname}{Assumption}

\newcommand{\Prb}{\mathbb{P}}
\newcommand{\E}{\mathbb{E}}
\newcommand{\PrbX}[1]{\Prb_{x_{0:#1},\tilde x_{0:#1}}}
\newcommand{\EX}[1]{\E_{x_{0:#1},\tilde x_{0:#1}}}
\newcommand{\EXzero}{\E_{x_{0},\tilde x_{0}}} 
\newcommand{\setX}{\mathbb{X}}
\newcommand{\reals}{\mathbb{R}}
\newcommand{\I}{\mathds{1}}
\newcommand{\tx}{\tilde{x}}
\newcommand{\tw}{\tilde{w}}
\newcommand{\ta}{\tilde{a}}

\newcommand{\tX}{\tilde{X}}

\newcommand{\transp}{\mathsf{T}}                      
\begin{document}

\def\spacingset#1{\renewcommand{\baselinestretch}%
{#1}\small\normalsize} \spacingset{1}


  \title{\bf Smoothing with Couplings of Conditional Particle Filters}
  \author{Pierre E. Jacob\thanks{
    The authors gratefully acknowledge the Swedish Foundation for Strategic Research (SSF)
via the projects \emph{Probabilistic Modeling and Inference for Machine Learning} (contract number: ICA16-0015) and ASSEMBLE (contract number: RIT15-0012), the Swedish Research Council (VR) via the projects 
\emph{Learning of Large-Scale Probabilistic Dynamical Models} (contract number: 2016-04278) and
\emph{NewLEADS - New Directions in Learning Dynamical Systems} (contract number: 621-2016-06079),
and the National Science Foundation through grant DMS-1712872.}
\hspace{.2cm}\\
  Department of Statistics, Harvard University\\
    Fredrik Lindsten and Thomas B. Sch\"on\\
    Department of Information Technology, Uppsala University}
  \maketitle

\begin{abstract}
In state space models, smoothing refers to the task of estimating a latent
stochastic process given noisy measurements related to the process. We propose an 
unbiased estimator of smoothing expectations. The lack-of-bias property has
methodological benefits: independent estimators can be generated in parallel,
and confidence intervals can be constructed from the central limit theorem to quantify the approximation error.
To design unbiased estimators, we combine a generic debiasing technique for Markov
chains with a Markov chain Monte Carlo algorithm for smoothing.  The resulting procedure is
widely applicable and we show in numerical experiments that the removal of the
bias comes at a manageable increase in variance.  We establish the validity of
the proposed estimators under mild assumptions. Numerical experiments are
provided on toy models, including a setting of highly-informative observations,
and a realistic Lotka-Volterra model with an intractable transition
density.
\end{abstract}

\noindent%

\spacingset{1.1}

\section{Introduction\label{sec:introduction}}

\subsection{Goal and content}

In state space models, the observations are treated as noisy measurements related to
an underlying latent stochastic process. The problem of smoothing refers to the estimation of trajectories of
the underlying process given the observations \citep{cappe:ryden:2004}. For finite
state spaces and linear Gaussian models, smoothing can be performed
exactly. In general models, numerical approximations are required, and
many state-of-the-art methods are based on particle methods
\citep{douc:moulines:2014,kantas2015particle}. Following this line of work,
we propose a new method for smoothing in general state space models.
Unlike existing methods, the proposed estimators are unbiased, which has direct
benefits for parallelization and for the construction of confidence intervals.

The proposed method combines recently proposed conditional particle filters
\citep{andrieu:doucet:holenstein:2010} with debiasing techniques for Markov
chains \citep{glynn2014exact}. 
Specifically, we show in Section \ref{sec:unbiasedsmoothing} how to remove the bias of 
estimators constructed with conditional particle filters,
in exchange for an increase of variance; this variance can then be controlled with tuning parameters,
and arbitrarily reduced by averaging over independent replicates.
The validity of the proposed approach
relies on the finiteness of the computational cost and of the variance of the proposed estimators, which we establish under mild conditions
in Section \ref{sec:newsmoother:theory}.
Methodological improvements are presented in Section \ref{sec:newsmoother:practical},
and comparisons with other smoothers in Section \ref{sec:comparison}.
Numerical experiments are provided in Section \ref{sec:numerics},
and Section \ref{sec:discussion} concludes.

\subsection{Smoothing in state space models \label{sec:intro:smoothing}}

The latent stochastic process $(x_{t})_{t\geq 0}$ takes values in $\mathbb{X}\subset
\mathbb{R}^{d_x}$, 
and the observations 
$(y_t)_{t\geq 1}$
are in $\mathbb{Y}\subset
\mathbb{R}^{d_y}$ for some $d_x,d_y \in\mathbb{N}$.  A model specifies an
initial distribution $m_0(dx_{0}|\theta)$ and a transition kernel
$f(dx_{t}| x_{t-1},\theta)$ for the latent process.
We will assume that we have access to deterministic functions $M$ and $F$,
and random variables $U_t$ for $t\geq 0$, such that $M(U_0,\theta)$ follows
$m_0(dx_0|\theta)$ and $F(x_{t-1},U_t,\theta)$ follows $f(dx_t|x_{t-1},\theta)$;
we refer to these as random function representations of the process \citep[see][]{diaconis1999iterated}.
Conditionally upon the latent process, the observations are independent
and their distribution is given by a measurement kernel $g(dy_{t}| x_{t},\theta)$. The model is
parameterized by $\theta\in\Theta\subset \mathbb{R}^{d_\theta}$, for $d_\theta\in\mathbb{N}$. 
Filtering consists in approximating the distribution $p(dx_{t}|
y_{1:t},\theta)$ for all times $t\geq 1$, whereas smoothing
refers to the approximation of $p(dx_{0:T}|y_{1:T},\theta)$ for
a fixed time horizon $T$, where
for $s,t\in\mathbb{N}$, we write $s:t$ for the set $\{s,\ldots,t\}$, and $v_{s:t}$
for the vector $(v_s,\ldots,v_t)$.
The parameter $\theta$ is hereafter fixed and removed from the notation,
as is usually done in the smoothing literature \citep[see Section 4 in][]{kantas2015particle};
we discuss unknown parameters in Section \ref{sec:discussion}.
Denote by $h$ a test function from $\mathbb{X}^{T+1}$ to $\mathbb{R}$, of
which we want to compute the expectation with respect to the smoothing distribution $\pi(dx_{0:T})=p(dx_{0:T}|y_{1:T})$; we write 
$\pi(h)$ for $\int_{\mathbb{X}^{T+1}} h(x_{0:T}) \pi(dx_{0:T})$. For instance, with $h:x_{0:T}\mapsto x_t$
where $t\in 0:T$, $\pi(h)$ is the smoothing expectation $\mathbb{E}[x_t|y_{1:T}]$.

Postponing a discussion on existing smoothing methods to Section \ref{sec:comparison},
we first describe the conditional particle filter \citep[CPF,][]{andrieu:doucet:holenstein:2010},
which is a variant of the particle filter \citep{doucet:defreitas:gordon:2001}.
Given a ``reference'' trajectory $X = x_{0:T}$, a CPF
generates a new trajectory $X^\prime = x_{0:T}^\prime$ as described in Algorithm \ref{alg:conditional-particle-filter},
which defines a Markov kernel on the space of trajectories; we will
write $x^\prime_{0:T} \sim \text{CPF}(x_{0:T},\cdot)$.
This Markov kernel leaves  $\pi$ invariant
and ergodic averages of the resulting chains consistently estimate integrals with respect to $\pi$,
under mild conditions \citep{andrieu:doucet:holenstein:2010,ChopinS:2015,LindstenDM:2015,andrieuvihola2013uniform,kuhlenschmidt2018stability,Lee2018ccbpf}.
We denote by $(X^{(n)})_{n\geq 0}$ a chain starting from a path $X^{(0)}$,
and iterating through $X^{(n)}\sim\text{CPF}(X^{(n-1)},\cdot)$ for $n\geq 1$. 

\begin{center}
    \begin{algorithm}
        {\small \textsf{At step $t=0$.}}
        \begin{enumerate}
            \item[1.1.] {\small \textsf{Draw $U_0^j$ and compute $x_{0}^{j} = M(U_0^j,\theta)$, for all $j=1,\ldots,N-1$, and $x_0^{N} = x_0$.}}
            \item[1.2.] {\small \textsf{Set $w_{0}^{j}=N^{-1}$, for $j=1,\ldots,N$.}}
        \end{enumerate}
        {\small \textsf{At step $t=1,\ldots,T$.}}
        \begin{enumerate}
            \item[2.1.] {\small \textsf{Draw ancestors $a_{t-1}^{1:N-1}\sim r(da^{1:N-1}| w_{t-1}^{1:N})$, and set $a_{t-1}^N = N$.}}
            \item[2.2.] {\small   \textsf{Draw $U_t^j$ and compute $x_{t}^{j} = F(x_{t-1}^{a_{t-1}^j}, U_t^j,\theta)$, for all $j=1,\ldots,N-1$, and $x_t^{N} = x_t$.}}
            \item[2.3.] {\small \textsf{Compute $w_{t}^{j} = g(y_{t}| x_{t}^{j},\theta)$, for all $j=1,\ldots,N$,
            and normalize the weights.}}
    \end{enumerate}
    {\small \textsf{After the final step.}}
    \begin{enumerate}
        \item[3.1.]  {\small \textsf{Draw $b_T$ from a discrete distribution on $1:N$, with probabilities $w^{1:N}_{T}$.}}
        \item[3.2.]  {\small \textsf{For $t= T-1,\ldots,0$, set $b_t = a_{t}^{b_{t+1}}$.}}
    \end{enumerate}
    {\small \textsf{Return $x_{0:T}^\prime = (x_0^{b_0}, \ldots, x_T^{b_T})$. }}
    \protect\caption{Conditional particle filter (CPF), given a trajectory $x_{0:T}$ and $\theta$. \label{alg:conditional-particle-filter}}
\end{algorithm}
\end{center}

In step 2.1. of Algorithm \ref{alg:conditional-particle-filter}, the resampling 
distribution $r(da^{1:N-1}|w^{1:N})$ refers to a distribution on $\{1,\ldots,N\}^{N-1}$
from which ``ancestors'' are drawn according to particle weights.
The resampling distribution is an algorithmic choice;
specific schemes for the conditional particle filter are described in \citet{ChopinS:2015}.
Here we will use multinomial resampling throughout.
In step 2.3., ``normalize the weights'' means dividing them by their sum.
Instead of bootstrap particle filters \citep{gordon:salmon:smith:1993},
where particles are propagated from the model transition,
more sophisticated filters can readily be used in the CPF procedure.
For instance, performance gains can be obtained with auxiliary particle filters \citep{pitt1999filtering,johansen2008note},
as illustrated in Section \ref{sec:numerics:hiddenar}.
In presenting algorithms we focus on bootstrap particle filters for simplicity.
When the transition density is tractable, extensions of the CPF include backward sampling
\citep{whiteleycommentonpmcmc,LindstenS:2013} and ancestor sampling
\citep{LindstenJS:2014}, which is beneficial in the proposed approach as illustrated in Section \ref{sec:numerics:hiddenar}. 
The complexity of a standard CPF update is of order $NT$,
and the memory requirements are of order $T + N\log N$ \citep{jacob2015path}.

The proposed method relies on CPF kernels but is different from Markov chain Monte Carlo (MCMC) estimators: it
involves independent copies of unbiased estimators of $\pi(h)$. Thus it will be amenable to parallel computation
and confidence intervals will be constructed in a different way than with standard MCMC output \citep[e.g. Chapter 7 in][]{gelman2010handbook};
see Section \ref{sec:comparison} for a comparison with existing smoothers.

\subsection{Debiasing Markov chains \label{sec:debiasing}}

We briefly recall the debiasing technique of \citet{glynn2014exact},
see also \citet{McLeish:2011,Rhee:Glynn:2012,vihola2015unbiased} and references therein.
Denote by $(X^{(n)})_{n\geq 0}$ and $(\tX^{(n)})_{n\geq 0}$ two Markov chains with invariant distribution $\pi$,
initialized from a distribution $\pi_0$.
Assume that, for all $n\geq 0$, $X^{(n)}$ and $\tX^{(n)}$ have the same marginal distribution,
and that $\lim_{n\to\infty} \mathbb{E}[h(X^{(n)})] = \pi(h)$.
Writing limit as a telescopic sum, and swapping
infinite sum and expectation, which will be justified later on, we obtain
\begin{align*}
    \pi(h) 
    &= \mathbb{E}[h(X^{(0)})] + \sum_{n=1}^\infty \mathbb{E}[h(X^{(n)}) - h(\tilde{X}^{(n-1)})]
    = \mathbb{E}[h(X^{(0)}) + \sum_{n=1}^\infty (h(X^{(n)}) - h(\tilde{X}^{(n-1)}))].
\end{align*}
Then, if it exists, the random variable $H_0 = h(X^{(0)}) + \sum_{n=1}^\infty (h(X^{(n)}) - h(\tilde{X}^{(n-1)}))$,
is an unbiased estimator of $\pi(h)$.  Furthermore, if the chains are coupled
in such a way that there exists a time~$\tau$, termed the \emph{meeting time}, such that
$X^{(n)}=\tX^{(n-1)}$ almost surely for all $n\geq \tau$, then $H_0$ can be computed as 
\begin{equation}
    H_0 = h(X^{(0)}) + \sum_{n=1}^{\tau - 1} (h(X^{(n)}) - h(\tilde{X}^{(n-1)})). \label{eq:RGestimator}
\end{equation}
We refer to $H_0$ as a Rhee--Glynn estimator. Given that the cost of producing $H_0$
increases with $\tau$, it will be worth keeping in mind
that we would prefer $\tau$ to take small values with
large probability. The main contribution of the present
article is to couple CPF chains and to use them
in a Rhee--Glynn estimation procedure. 
Section \ref{sec:newsmoother:theory} provides guarantees on the cost and the variance
of $H_0$ under mild conditions, and Section \ref{sec:newsmoother:practical}
contains alternative estimators with reduced variance and practical considerations.

\section{Unbiased smoothing \label{sec:unbiasedsmoothing}}

\subsection{Coupled conditional particle filters \label{sec:ccpf}}

Our goal is to couple CPF  
chains $(X^{(n)})_{n\geq 0}$ and $(\tX^{(n)})_{n\geq 0}$
such that the meeting time has finite expectation, in order to enable a Rhee--Glynn estimator for smoothing.
A coupled conditional particle filter (CCPF) is 
a Markov kernel on the space of pairs of trajectories,
such that $(X^\prime,\tX^\prime)\sim \text{CCPF}((X,\tX), \cdot)$
implies that $X^\prime\sim \text{CPF}(X, \cdot)$
and $\tX^\prime \sim \text{CPF}(\tX, \cdot)$.

Algorithm \ref{alg:coupled-conditional-particle-filter}
describes CCPF in pseudo-code, conditional upon $X = x_{0:T}$ and $\tX = \tx_{0:T}$.
Two particle systems are initialized and propagated using common random numbers.
The resampling steps and the selection of trajectories at the final step are performed jointly
using couplings of discrete distributions.
To complete the description of the CCPF procedure,
we thus need to specify these couplings (for steps 2.1. and 3.1. in Algorithm \ref{alg:coupled-conditional-particle-filter}). With the Rhee--Glynn estimation procedure
in mind, we aim at achieving large meeting probabilities $\mathbb{P}(X^\prime = \tX^\prime | X,\tX)$,
so as to incur short meeting times on average.

\noindent \begin{center}
    \begin{algorithm}
        {\small \textsf{At step $t=0$.}}
        \begin{enumerate}
            \item[1.1.]{\small \textsf{Draw $U_{0}^{j}$, compute $x_{0}^{j} = M(U_0^j,\theta)$ and $\tx_{0}^{j} = M(U_0^j,\theta)$ for $j = 1,\ldots,N-1$.}}
            \item[1.2.]{\small \textsf{Set  $x_0^{N} = x_0$ and $\tx_0^N=\tx_0$.}}
            \item[1.3.]{\small \textsf{Set $w_{0}^{j}=N^{-1}$ and $\tw_{0}^{j}=N^{-1}$, for $j=1,\ldots,N$.}}
        \end{enumerate}
        {\small \textsf{At step $t=1,\ldots,T$.}}
        \begin{enumerate}
            \item[2.1.] {\small \textsf{Draw ancestors $a_{t-1}^{1:N}$ and $\ta_{t-1}^{1:N}$
                from a coupling of $r(da^{1:N-1}|w_{t-1}^{1:N})$ and $r(da^{1:N-1}|\tw_{t-1}^{1:N})$, and set $a_{t-1}^N = N$ and $\ta_{t-1}^N = N$.}}
            \item[2.2.] {\small \textsf{Draw $U_{t}^j$, and compute $x_{t}^{j} = F(x_{t-1}^{a_{t-1}^{j}}, U_{t}^j, \theta)$ and $\tx_{t}^{j} = F(\tx_{t-1}^{\tilde{a}_{t-1}^{j}}, U_{t}^j, \theta)$, for all $j=1,\ldots,N-1$. Set  $x_t^{N} = x_t$ and $\tx_t^{N} = \tx_t$.}}
            \item[2.3.] {\small \textsf{Compute $w_{t}^{j} = g(y_{t}| x_{t}^{j},\theta)$ and
                $\tw_{t}^{j} =  g(y_{t}| \tx_{t}^{j},\theta)$, for all $j=1,\ldots,N$,
            and normalize the weights.}}
    \end{enumerate}
    {\small \textsf{After the final step.}}
    \begin{enumerate}
        \item[3.1.] {\small \textsf{Draw $(b_{T}, \tilde{b}_{T})$ from a coupling of two distributions on $1:N$, with probabilities $w_{T}^{1:N}$ and $\tw_{T}^{1:N}$ respectively.}}
        \item[3.2.] {\small \textsf{For $t= T-1,\ldots,0$, set $b_t = a_{t}^{b_{t+1}}$ and $\tilde{b}_t = \ta_{t}^{\tilde{b}_{t+1}}$.}}
    \end{enumerate}
    {\small \textsf{Return $x_{0:T}^\prime = (x_0^{b_0}, \ldots, x_T^{b_T})$ and $\tx_{0:T}^\prime = (\tx_0^{\tilde{b}_0}, \ldots, \tx_T^{\tilde{b}_T})$. }}
    \protect\caption{Coupled conditional particle filter (CCPF), given trajectories $x_{0:T}$ and $\tx_{0:T}$. \label{alg:coupled-conditional-particle-filter}}
\end{algorithm}

\par\end{center}

\subsection{Coupled resampling \label{sec:couplingparticlesystems}}

The temporal index $t$ is momentarily removed from the notation: the task is that of sampling 
pairs $(a,\ta)$ such that
$\mathbb{P}(a=j)=w^{j}$ and $\mathbb{P}(\ta=j)=\tw^{j}$ for all $j\in 1:N$;
this is a sufficient condition for CPF kernels
to leave $\pi$ invariant \citep{andrieu:doucet:holenstein:2010}.

A joint distribution on $\{1,\ldots,N\}^{2}$ is characterized by a matrix $P$ with non-negative entries
$P^{ij}$, for $i,j\in\{ 1,\ldots,N\}$, that sum to one. The value $P^{ij}$ represents the
probability of the event $(a,\ta) = (i,j)$. We consider the set
$\mathcal{J}(w,\tw)$ of matrices $P$ such that 
$P\mathds{1}=w$ and $P^\transp\mathds{1}=\tw$, where $\mathds{1}$ denotes a
column vector of $N$ ones, $w = w^{1:N}$ and $\tw = \tw^{1:N}$. Matrices $P\in \mathcal{J}(w,\tw)$ are such that
$\mathbb{P}(a=j)=w^{j}$ and $\mathbb{P}(\ta=j)=\tw^{j}$ for $j\in 1:N$. 

Any choice of probability matrix $P\in\mathcal{J}(w,\tw)$, and of a way of sampling $(a,\ta)\sim P$, leads to a \emph{coupled}  resampling scheme.
In order to keep the complexity of sampling $N$ pairs from $P$ linear in $N$,
we focus on a particular choice.
Other choices of coupled resampling schemes are given in \citet{deligiannidis2015correlated,jacob2016coupling,sen2018coupling},
following earlier works such as \citet{pitt2002smooth,lee2008towards}. 

We consider the \emph{index-coupled} resampling scheme, used by
\citet{ChopinS:2015}  in their theoretical analysis of the CPF,
and by \citet{jasra2015multilevel} in a multilevel Monte Carlo context,
see also Section 2.4 in \citet{jacob2016coupling}.
The scheme amounts to a maximal coupling of discrete distributions
on $\{1,\ldots,N\}$ with probabilities $w^{1:N}$ and $\tw^{1:N}$,  respectively.
This coupling maximizes the probability of the event $\{a = \tilde{a}\}$ under the marginal constraints.
How to sample from a maximal coupling of discrete distributions is described e.g. in \citet{lindvall2002lectures}.
The scheme is intuitive at the initial step of the CCPF, when $x_0^j = \tx_0^j$ for all $j=1,\ldots,N-1$:
one would want pairs of ancestors $(a_0,\ta_0)$ to be such that $a_0 = \ta_0$,
so that pairs of resampled particles remain identical.
At later steps, the number of identical pairs across both particle systems might be small, or even null.
In any case, at step 2.2. of Algorithm \ref{alg:coupled-conditional-particle-filter}, 
the same random number $U_{t}^j$ is used to compute $x^j_{t}$ and $\tx^j_{t}$
from their ancestors. If $a_{t-1}^j = \ta_{t-1}^j$, we select ancestor particles that were,
themselves, computed with common random numbers at the previous step, and we give them common random numbers
again.  Thus this scheme maximizes the number of consecutive steps at which common random numbers are used
to propagate each pair of particles.

We now discuss why propagating pairs of particles with common random numbers might be desirable. 
Under assumptions on the random function representation of the latent process,
using common random numbers to propagate pairs of particles 
results in the particles contracting. 
For instance, in an auto-regressive model where $F(x,U,\theta) = \theta x + U$,
where $\theta \in (-1,1)$ and $U$ is the innovation term,
we have $|F(x,U,\theta) - F(\tx,U,\theta)| = |\theta| |x-\tx|$, thus a pair of particles 
propagated with common variables $U$ contracts at a geometric rate.
We can formulate
assumptions directly on the function  $x\mapsto \mathbb{E}_U[F(x,U,\theta)]$,
such as Lipschitz conditions with respect to $x$, after having integrated $U$ out,
for fixed $\theta$.
Discussions on these assumptions can be found in \citet{diaconis1999iterated},
and an alternative method that would not require them is mentioned in 
Section \ref{sec:discussion}.

\subsection{Rhee--Glynn smoothing estimator \label{sec:rgsmoothing}}

We now put together the Rhee--Glynn estimator of Section \ref{sec:debiasing}
with the CCPF algorithm of Section \ref{sec:ccpf}. In passing we generalize
the Rhee--Glynn estimator slightly by starting the telescopic sum at index $k\geq 0$
instead of zero, and denote it by $H_k$; $k$ becomes a tuning parameter, discussed in Section \ref{sec:newsmoother:practical}.
The procedure is fully described in Algorithm \ref{alg:rheeglynnsmoother};
CPF and CCPF refer to Algorithms \ref{alg:conditional-particle-filter} and \ref{alg:coupled-conditional-particle-filter} respectively.

By convention the sum from $k+1$ to $\tau-1$ in
the definition of $H_k$ is set to zero whenever $k+1>\tau-1$. Thus 
the estimator $H_k$ is equal to $h(X^{(k)})$ on the event $\{k+1>\tau-1\}$.
Recall that $h(X^{(k)})$ is in general a biased estimator of $\pi(h)$,
since there is no guarantee that a CPF chain reaches stationarity within $k$ iterations.
Thus the term $\sum_{n=k+1}^{\tau - 1}(h(X^{(n)}) - h(\tX^{(n-1)}))$ acts as a bias correction.

\begin{algorithm}
    \begin{enumerate}
        \item[1.] {\small \textsf{Draw $X^{(0)}\sim\pi_0$, $\tX^{(0)}\sim\pi_0$ and draw $X^{(1)} \sim \text{CPF}(X^{(0)}, \cdot)$.}}
        \item[2.] {\small\textsf{Set $n = 1$. While $n < \max(k,\tau)$, where $\tau = \inf\{n\geq 1: X^{(n)} = \tX^{(n-1)}\}$,}}
            \begin{enumerate}
                \item[2.1.] {\small \textsf{Draw $(X^{(n+1)},\tilde{X}^{(n)}) \sim \text{CCPF}((X^{(n)},\tilde{X}^{(n-1)}), \cdot)$.}}
                \item[2.2.] {\small \textsf{Set $n \leftarrow n + 1$.}}
            \end{enumerate}
        \item[3.] {\small \textsf{Return $H_k = h(X^{(k)}) + \sum_{n = k+1}^{\tau-1} (h(X^{(n)}) - h(\tX^{(n-1)}))$.}}
    \end{enumerate}
\protect\caption{Rhee--Glynn smoothing estimator, with initial $\pi_0$ and tuning parameter $k$. \label{alg:rheeglynnsmoother}}
\end{algorithm}

At step 1. of Algorithm \ref{alg:rheeglynnsmoother}, the paths $X^{(0)}$ and $\tX^{(0)}$
can be sampled independently or not from $\pi_0$. In the experiments we will
initialize chains independently and $\pi_0$ will refer to the distribution
of a path randomly chosen among the trajectories of a 
particle filter.

\section{Theoretical properties\label{sec:newsmoother:theory}} 

We give three sufficient conditions for the validity of Rhee--Glynn smoothing estimators.

\begin{assumption}
\label{assumption:upperbound}  
    The measurement density of the model is bounded from above:
    there exists $\bar{g} < \infty$ such that, for all $y\in \mathbb{Y}$ and $x\in\mathbb{X}$, $g(y | x) \leq \bar{g}$.
\end{assumption}

\begin{assumption}
    \label{assumption:couplingmatrix}
    The resampling probability matrix $P$, with rows summing to $w^{1:N}$ and columns summing to $\tw^{1:N}$, is such that,
    for all $i\in \{1,\ldots,N\}$, $P^{ii} \geq w^i \tw^i$.
    Furthermore, if $w^{1:N} = \tw^{1:N}$, then $P$ is a diagonal matrix with entries given by $w^{1:N}$.
\end{assumption}

\begin{assumption}
\label{assumption:mixing}  
Let $(X^{(n)})_{n \geq 0}$ be a Markov chain generated by the conditional particle filter and started from $\pi_0$,
and $h$ a test function of interest. Then $\mathbb{E}\left[h(X^{(n)})\right] \xrightarrow[n\to \infty]{} \pi(h)$.
Furthermore, there exists $\delta > 0$, $n_0 < \infty$ and $C<\infty$ such that,
for all $n\geq n_0$, $\mathbb{E}\left[h(X^{(n)})^{2+\delta}\right]\leq C$.
\end{assumption}

The first assumption is satisfied for wide classes of models where the
measurements are assumed to be some transformation of the latent process with
added noise. However, it would not be satisfied for instance in stochastic
volatility models where it is often assumed that $Y|X=x\sim \mathcal{N}(0,
\exp(x)^2)$ or variants thereof \citep[e.g.][]{fulop2013efficient}. There, the
measurement density would diverge when $y$ is exactly zero and $x\to -\infty$. A similar assumption
is discussed in Section 3 of \citet{whiteley2013stability}.
One can readily check that the second assumption always holds for the index-coupled resampling scheme.
The third assumption relates to the validity of MCMC estimators generated by the CPF algorithm,
addressed under general assumptions in
\citet{ChopinS:2015,LindstenDM:2015,andrieuvihola2013uniform}. 

Our main result states that the proposed estimator is unbiased, has a finite variance,
and that the meeting time $\tau$ has tail probabilities bounded by those of a geometric variable, which implies in particular
that the estimator has a finite expected cost. 

\begin{thm}
    Under Assumptions \ref{assumption:upperbound} and \ref{assumption:couplingmatrix},
    for any initial distribution $\pi_0$, any number of particles $N\geq 2$ and time horizon $T\geq 1$,
    there exists $\varepsilon>0$, which might depend on $N$ and $T$, such that 
    for all $n\geq 2$,
    \[\mathbb{P}(\tau > n) \leq (1-\varepsilon)^{n-1},\]
    and therefore $\mathbb{E}[\tau]<\infty$. 
    Under the additional Assumption \ref{assumption:mixing}, the Rhee--Glynn smoothing estimator $H_k$ of Algorithm \ref{alg:rheeglynnsmoother}
    is such that, for any $k\geq 0$, $\mathbb{E}[H_k] = \pi(h)$ and $\mathbb{V}[H_k] < \infty$.
    \label{thm:finitevariance}
\end{thm}

The proof is in Appendices \ref{sec:proof:intermed} and \ref{sec:proof:unbiased}.
Some aspects of the proof, not specific to the smoothing setting, 
are similar to the proofs of Theorem 1 in \citet{rhee:phd}, Theorem 2.1 in \citet{McLeish:2011},
Theorem 7 in \citet{vihola2015unbiased}, and results in \citet{glynn2014exact}.
It is provided
in univariate notation  but the Rhee--Glynn
smoother can estimate multivariate smoothing functionals,
in which case the theorem applies component-wise.

\section{Improvements and tuning \label{sec:newsmoother:practical}} 

Since $H_\ell$ is unbiased for all $\ell\geq 0$, we can compute
$H_\ell$ for various values of $\ell$ between two integers $k\leq m$, and average these estimators to obtain  $H_{k:m}$ defined as 
\begin{align}\label{eq:timeaverage}
H_{k:m} & = \frac{1}{m-k+1}\sum_{n = k}^m \{h(X^{(n)})  + \sum_{\ell = n + 1}^{\tau - 1} (h(X^{(\ell)}) - h(\tX^{(\ell-1)}))\} \nonumber \\
   &= \frac{1}{m-k+1}\sum_{n = k}^m h(X^{(n)}) +  \sum_{n =k + 1}^{\tau - 1} \frac{\min(m-k+1, n-k)}{m-k+1} (h(X^{(n)}) - h(\tX^{(n-1)})).
\end{align}
The term $(m-k+1)^{-1} \sum_{n = k}^m h(X^{(n)})$ 
is a standard ergodic average of a CPF chain, after $m$ iterations and discarding the first $k-1$ steps as burn-in.
It is a biased estimator of $\pi(h)$ in general since $\pi_0$ is different from $\pi$.
The other term acts as a bias correction. 
On the event $\tau - 1< k+1$ the correction term is equal to zero. 

As $k$ increases the bias of the term $(m-k+1)^{-1} \sum_{n = k}^m h(X^{(n)})$ decreases.
The variance inflation of the Rhee--Glynn estimator decreases too, since the correction term is equal to zero
with increasing probability. On the other hand, it can be wasteful to set $k$ to an overly large value,
in the same way that it is wasteful to discard too many iterations as burn-in when computing MCMC estimators.
In practice we propose to choose $k$ according to the distribution of $\tau$, which can be sampled from
exactly by running Algorithm \ref{alg:rheeglynnsmoother}, as illustrated in the numerical experiments
of Section \ref{sec:numerics}. Conditional upon a choice of $k$, by analogy with MCMC estimators
we can set $m$ to a multiple of $k$, such as $2k$ or $5k$. Indeed the proportion of discarded iterations
is approximately $k/m$, and it appears desirable to keep this proportion low. 
We stress that the proposed estimators are unbiased and with a finite variance for any choice of $k$ and $m$;
tuning $k$ and $m$ only impacts variance and cost. 

For a given choice of $k$ and $m$, the estimator $H_{k:m}$ can be sampled $R$ times
independently in parallel. We denote the independent copies by $H_{k:m}^{(r)}$ for $r\in 1:R$.
The smoothing expectation of interest $\pi(h)$ can then be approximated by $\bar{H}_{k:m}^R = R^{-1}\sum_{r=1}^R H_{k:m}^{(r)}$,
with a variance that decreases linearly with $R$. From the central limit theorem
the confidence interval $[\bar{H}_{k:m}^R + z_{\alpha/2} \hat{\sigma}^R/\sqrt{R}, \bar{H}_{k:m}^R + z_{1-\alpha/2} \hat{\sigma}^R/\sqrt{R}]$,
where $\hat{\sigma}^R$ is the empirical standard deviation of $(H_{k:m}^{(r)})_{r=1}^R$ and $z_a$ is the $a$-th quantile of a standard Normal distribution,
has $1-\alpha$ asymptotic coverage as $R\to \infty$. The central limit theorem is applicable as a consequence of Theorem \ref{thm:finitevariance}.

The variance of the proposed estimator can be further reduced by Rao--Blackwellization.  
In Eq.~\eqref{eq:timeaverage}, 
the random variable $h(X^{(n)})$ is obtained by applying the test function $h$ of interest to 
a trajectory drawn among $N$ trajectories, denoted by say $x_{0:T}^k$ for $k=1,\ldots,N$, with probabilities $w_T^{1:N}$;
see step 3 in Algorithms \ref{alg:conditional-particle-filter} and \ref{alg:coupled-conditional-particle-filter}.
Thus the random variable $\sum_{k=1}^N w_T^{k}h(x_{0:T}^{k})$ is the conditional expectation
of $h(X^{(n)})$ given the trajectories and $w_T^{1:N}$, which
has the same expectation as $h(X^{(n)})$. Thus any term $h(X^{(n)})$ or
$h(\tX^{(n)})$ in $H_{k:m}$ can be replaced by similar conditional expectations.
This enables the use of all the paths generated by the CPF and CCPF kernels, and not only the selected ones.

As in other particle methods the choice of the number of particles $N$ is important.
Here, the estimator $\bar{H}_{k:m}^R$ is consistent as $R\to \infty$ for any $N\geq 2$, but $N$ plays a role both on the cost and of the variance of each $H^{(r)}_{k:m}$.
We can generate unbiased estimators for different values of $N$ and compare their costs and variances in preliminary runs.
The scaling of $N$ with the time horizon $T$ is explored numerically in Section \ref{sec:numerics:hiddenar}.
If possible, one can also employ other algorithms than the bootstrap particle filter,
as illustrated in Section \ref{sec:numerics:hiddenar} with the auxiliary particle filter.

\section{Comparison with existing smoothers \label{sec:comparison}}

The proposed method combines elements from both particle smoothers and MCMC methods,
but does not belong to either category.
We summarize advantages and drawbacks below, after having discussed
the cost of the proposed estimators.

Each estimator $H_{k:m}$ requires two draws from $\pi_0$, here taken as the distribution of a trajectory selected from a particle filter
with $N$ particles. Then, the estimator as described in Algorithm \ref{alg:rheeglynnsmoother} requires a draw from the CPF kernel,
$\tau-1$ draws from the CCPF kernel, and finally $m-\tau$ draws of the CPF kernel on the events $\{m>\tau\}$.
The cost of a particle filter and of an iteration of CPF is usually dominated
by the propagation of $N$ particles and the evaluation of their weights. The cost of an iteration of CCPF
is approximately twice larger. 
Overall the cost of $H_{k:m}$ is thus of order $C(\tau,m,N) = N\times (3+2(\tau-1)+\max(0,m-\tau))$, for fixed $T$.
The finiteness of the expected cost $\mathbb{E}[C(\tau,m,N)]$ is a consequence of Theorem~\ref{thm:finitevariance}. The average $\bar{H}_{k:m}^R$ satisfies a central limit theorem
parametrized by the number of estimators $R$, as discussed in Section \ref{sec:newsmoother:practical}; however, since the cost of $H_{k:m}$ is random, it might be 
more relevant to consider central limit theorems parametrized by computational cost,  as in \citet{glynn1992asymptotic}.
The asymptotic inefficiency of the proposed estimators can be defined as $\mathbb{E}[C(\tau,m,N)]\times\mathbb{V}[H_{k:m}]$,
which can be approximated with independent copies of $H_{k:m}$ and $\tau$, obtained by running Algorithm \ref{alg:rheeglynnsmoother}.

State-of-the-art particle smoothers include fixed-lag approximations
\citep{kitagawa2001monte,cappe:ryden:2004,olsson2008sequential}, forward
filtering backward smoothers
\citep{GodsillDW:2004,del2010forward,douc2011sequential,taghavi2013adaptive},
and smoothers based on the two-filter formula
\citep{briers2010smoothing,kantas2015particle}. These particle methods provide
consistent approximations as $N\to\infty$, with associated mean squared error decreasing as $1/N$ \citep[Section 4.4
of][]{kantas2015particle}; except for fixed-lag approximations for which 
some bias remains. The cost is typically of order $N$ with efficient
implementations described in \citet{fearnheadwyncolltawn2010,kantas2015particle,olsson2017efficient}, and is linear in $T$ for fixed $N$.
Parallelization over the $N$ particles is mostly feasible, the main limitation coming from the resampling step
\citep{murray2015parallel,lee2015forest,whiteley2016role,paige2014asynchronous,murray2016anytime}.
The memory cost of particle filters is of order $N$, or $N\log N$ if trajectories are kept \citep{jacob2015path}, see also \citet{Koskela2018}.
Assessing the accuracy of particle approximations from a single run of these methods remains a major challenge; see
\citet{lee2015variance,olsson2017numerically} for recent breakthroughs. 
Furthermore, we will see in Section \ref{sec:numerics:unlikely} that the bias of particle smoothers cannot always be safely ignored.
On the other hand, we will see in Section \ref{sec:numerics:pz} that the variance of particle smoothers can be smaller than that of the proposed estimators,
for a given computational cost. Thus, in terms of mean squared error per unit of computational cost, the proposed
method is not expected to provide benefits. 

The main advantage of the proposed method over particle smoothers lies in the construction of confidence intervals,
and the possibility of parallelizing over independent runs as opposed to interacting particles.
Additionally, a user of particle smoothers who would want
more precise results would increase the number
of particles $N$, if enough memory is available, discarding previous runs. On the other hand, the proposed estimator
$\bar{H}_{k:m}^R$ can be refined to arbitrary precision by drawing more independent copies of
$H_{k:m}$, for a constant memory requirement.

Other popular smoothers belong to the family of MCMC methods. Early examples include Gibbs
samplers, updating components of the latent process conditionally on other
components and on the observations \citep[e.g.][]{carter1994gibbs}.  The CPF
kernel described in Section~\ref{sec:intro:smoothing} can be used in the standard 
MCMC way, averaging over as many iterations as possible
\citep{andrieu:doucet:holenstein:2010}.  The bias of MCMC estimators after a
finite number of iterations is hard to assess, which makes the choice of
burn-in period difficult.  Asymptotically valid confidence intervals can be
produced in various ways, for instance using the CODA package
\citep{plummer2006coda}; see also \citet{vats2018strong}. On the other hand, parallelization over the iterations is
intrinsically challenging with MCMC methods \citep{rosenthal2000parallel}. 

Therefore the proposed estimators have some advantages over existing methods, the main drawback being a potential 
increase in mean squared error for a given (serial) computational budget, as illustrated in the numerical experiments.

\section{Numerical experiments\label{sec:numerics}}

We illustrate the tuning of the proposed estimators, their advantages and their drawbacks 
through numerical experiments. All estimators of this section employ the Rao--Blackwellization technique
described in Section \ref{sec:newsmoother:practical}, and multinomial resampling
is used within all filters.

\subsection{Hidden auto-regressive model\label{sec:numerics:hiddenar}}

Our first example illustrates the proposed method, 
the impact of the number of
particles $N$ and that of the time horizon $T$, and the benefits of auxiliary particle filters.
We consider a linear Gaussian model, with  $x_{0}\sim\mathcal{N}\left(0,1\right)$ and $x_{t}=\eta
x_{t-1}+\mathcal{N}\left(0,1\right)$ for all $t \geq 1$, with $\eta=0.9$.
We assume that $y_{t}\sim\mathcal{N}\left(x_{t},1\right)$ for all $t \geq 1$.

\begin{figure}
\begin{centering}
\subfloat[\label{fig:ar1:meetings} Meeting times.]{\begin{centering}
    \includegraphics[width=0.325\textwidth]{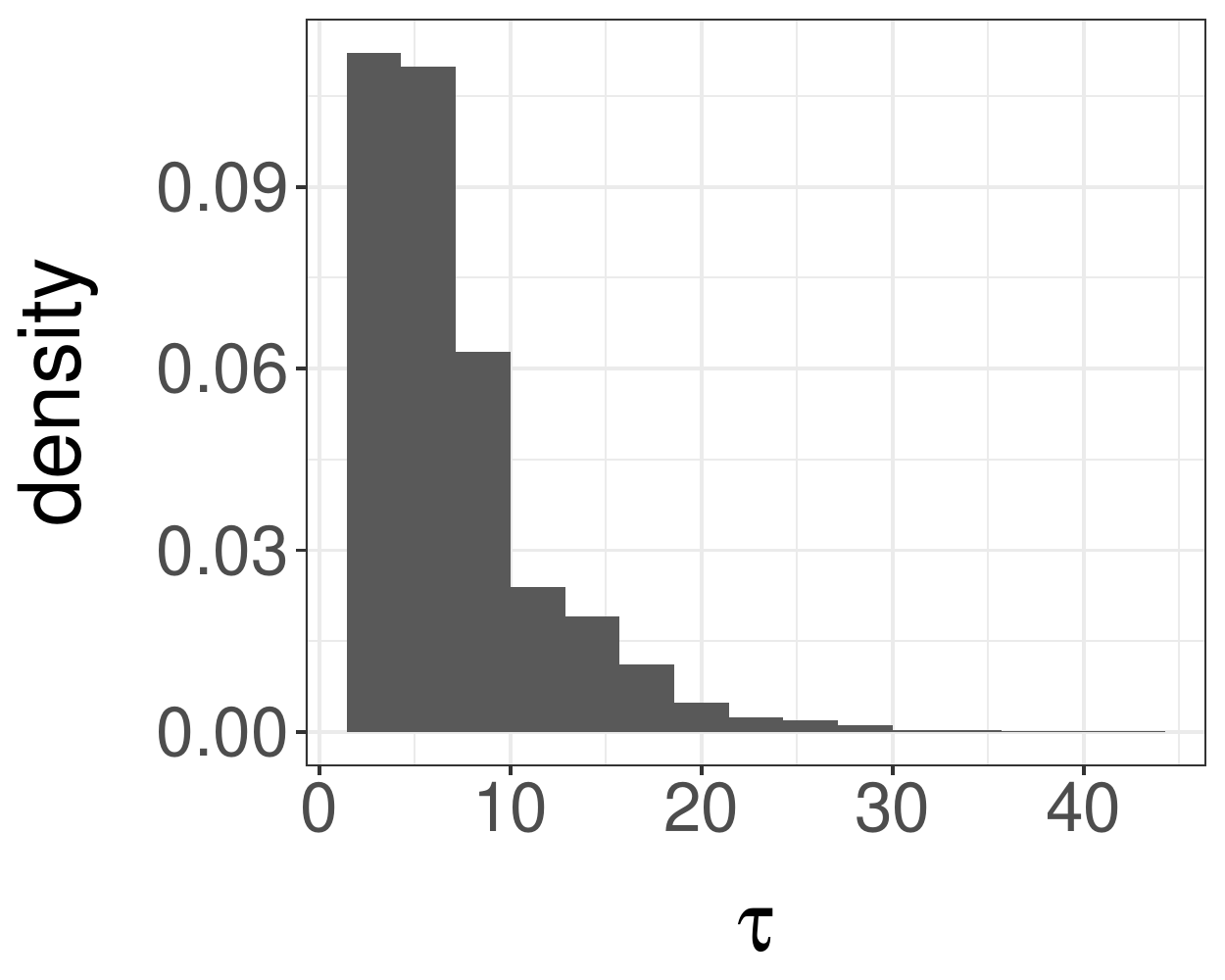}
\par\end{centering}
}
\subfloat[\label{fig:ar1:smoothingmeans} Smoothing means: 95\%  confidence intervals as error bars in black, and exact values connected by a red line. ]{\begin{centering}
    \includegraphics[width=0.65\textwidth]{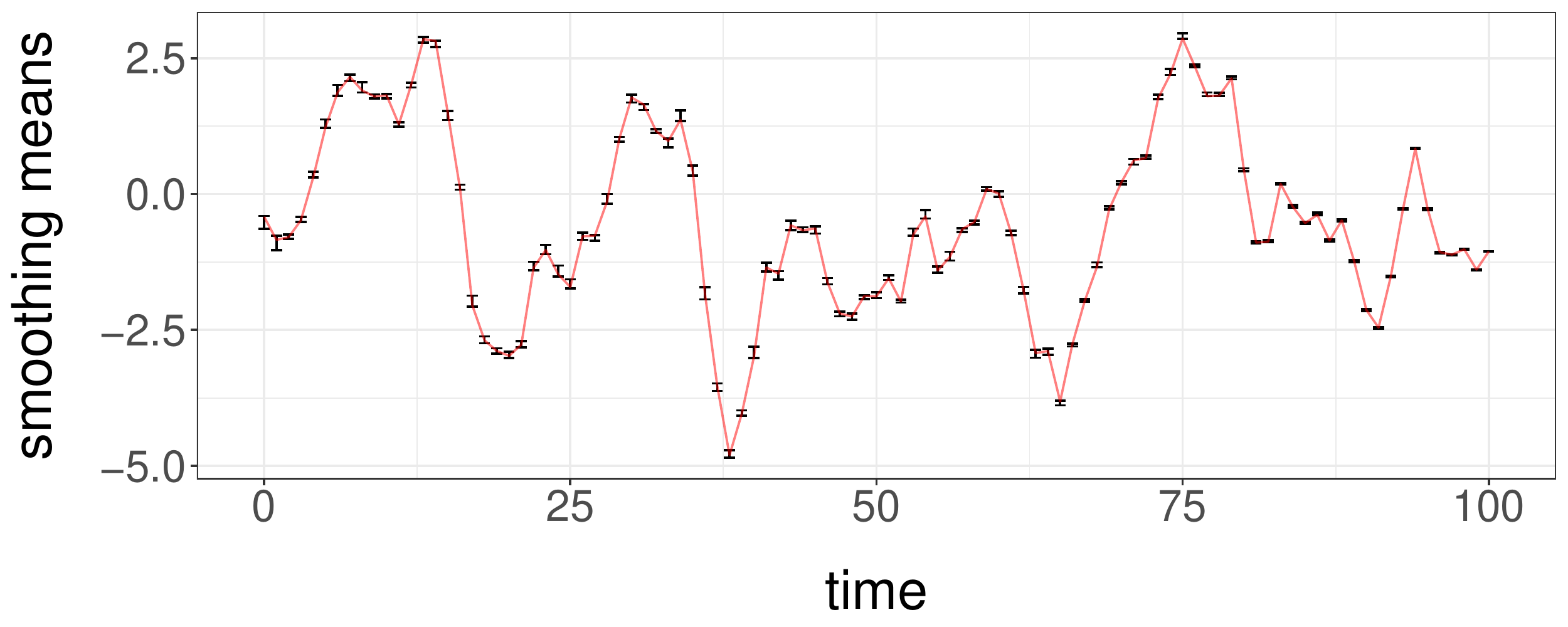}
\par\end{centering}
}
\par\end{centering}
\caption{\label{fig:ar1} Experiments with the auto-regressive model of Section \ref{sec:numerics:hiddenar}, with $T=100$ observations.
Here the CPF kernel employs $N=256$ particles and ancestor sampling. The distribution of the meeting times of coupled chains
is shown on the left, and error bars for the estimation of smoothing means, obtained with $k=10$ and $m=20$ over $R=100$ estimators, are shown on the right.}
\end{figure}

We first generate $T = 100$ observations from the model, and consider the task
of estimating all smoothing means, which corresponds to the test function $h:
x_{0:T}\mapsto x_{0:T}$.  With CPF kernels using bootstrap particle filters,
with $N = 256$ particles and ancestor sampling \citep{LindstenJS:2014}, we draw
meeting times $\tau$ independently, and represent a histogram of them in Figure
\ref{fig:ar1:meetings}.  Based on these meeting times, we can choose $k$ as a
large quantile of the meeting times, for instance $k = 10$, and $m$ as a
multiple of $k$, for instance $m = 2k = 20$.  For this choice, we find the
average compute cost of each estimator to approximately equal that of a
particle filter with $28\times 256$ particles, with a memory usage equivalent
to $2\times 256$ particles.  How many of these estimators can be produced in a
given wall-clock time depends on available hardware.  With $R=100$ independent
estimators, we obtain $95\%$ confidence intervals indicated by black error bars
in Figure \ref{fig:ar1:smoothingmeans}. The true smoothing means, obtained by
Kalman smoothing, are indicated by a line. 

The method is valid for all $N$, which prompts the question of the optimal
choice of $N$.  Intuitively, larger values of $N$ lead to smaller meeting
times. However, the meeting time cannot be less than $2$ by definition, which
leads to a trade-off. We verify this intuition by numerical simulations with
$1,000$ independent runs.  For $N=16$, $N=128$, $N=256$, $N=512$ and $N=1,024$, we find
average meeting times of $97$, $15$, $7$, $4$ and $3$ respectively. After adjusting
for the different numbers of particles, the expected cost of
obtaining a meeting is approximately equivalent with $N=16$ and $N=512$, but
more expensive for $N=1,024$. In practice, for
specific integrals of interest, one can approximate the cost and the variance
of the proposed estimators for various values of $N$, $k$ and $m$ using
independent runs, and use the most favorable configuration in subsequent,
larger experiments.

Next we investigate the effect of the time horizon $T$. We expect the performance of the CPF
kernel to decay as $T$ increases for a fixed $N$. We compensate by increasing $N$ linearly with $T$. 
Table \ref{table:effecthorizon} reports the average meeting times
obtained from $R=500$ independent runs.
We see that the average meeting times are approximately constant or slightly decreasing over $T$,  implying
that the linear scaling of $N$ with $T$ is appropriate or even conservative, in agreement with the literature \citep[e.g.][]{huggins2015sequential}. The table
contains the average meeting times obtained with and without ancestor sampling \citep{LindstenJS:2014};
we observe significant reductions of average meeting times with ancestor sampling, but it requires
tractable transition densities. 
Finally, for the present model we can employ an auxiliary particle filter,
in which particles are propagated conditionally on the next observation. Table \ref{table:effecthorizon}
shows a significant
reduction in expected meeting time. The combination of auxiliary particle filter
and ancestor sampling naturally leads to the smallest expected meeting times.

\input{tablear1horizon}

\subsection{A hidden auto-regressive model with an unlikely observation}\label{sec:numerics:unlikely}%
We now illustrate the benefits of the proposed estimators  in an example
taken from \citet{ruiz2016particle}
where particle filters exhibit a significant bias. The latent process
is defined as $x_{0}\sim\mathcal{N}\left(0,0.1^{2}\right)$ and $x_{t}=\eta
x_{t-1}+\mathcal{N}\left(0,0.1^{2}\right)$; we take $\eta=0.9$
and consider $T=10$ time steps. The process is observed only at
time $T=10$, where $y_{T}=1$ and we assume
$y_{T}\sim\mathcal{N}\left(x_{T},0.1^{2}\right)$. The
observation $y_{T}$ is unlikely under the model.
Therefore the filtering distributions and the smoothing distributions have
little overlap, particularly for times $t$ close to $T$. This toy model
is a stylized example of settings with highly-informative observations \citep{ruiz2016particle,del2015sequential}. 

\begin{figure}
\begin{centering}
\subfloat[\label{fig:unlikely:pf} Distributions of smoothing estimates produced by particle filters.]{\begin{centering}
    \includegraphics[width=0.4\textwidth]{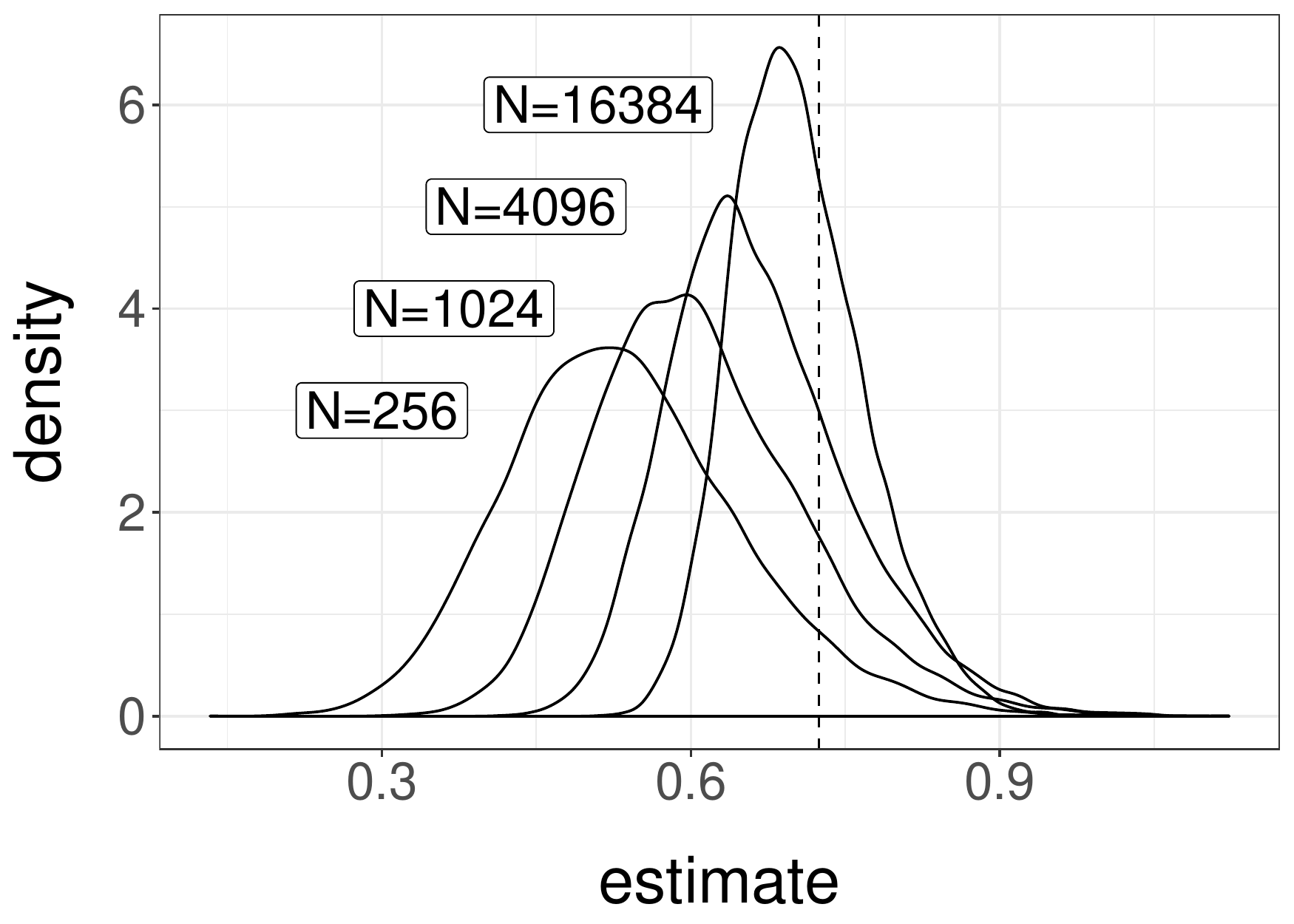}
\par\end{centering}
}\hspace*{1cm}
\subfloat[\label{fig:unlikely:rg} 95\% confidence intervals produced by the proposed method. ]{\begin{centering}
    \includegraphics[width=0.4\textwidth]{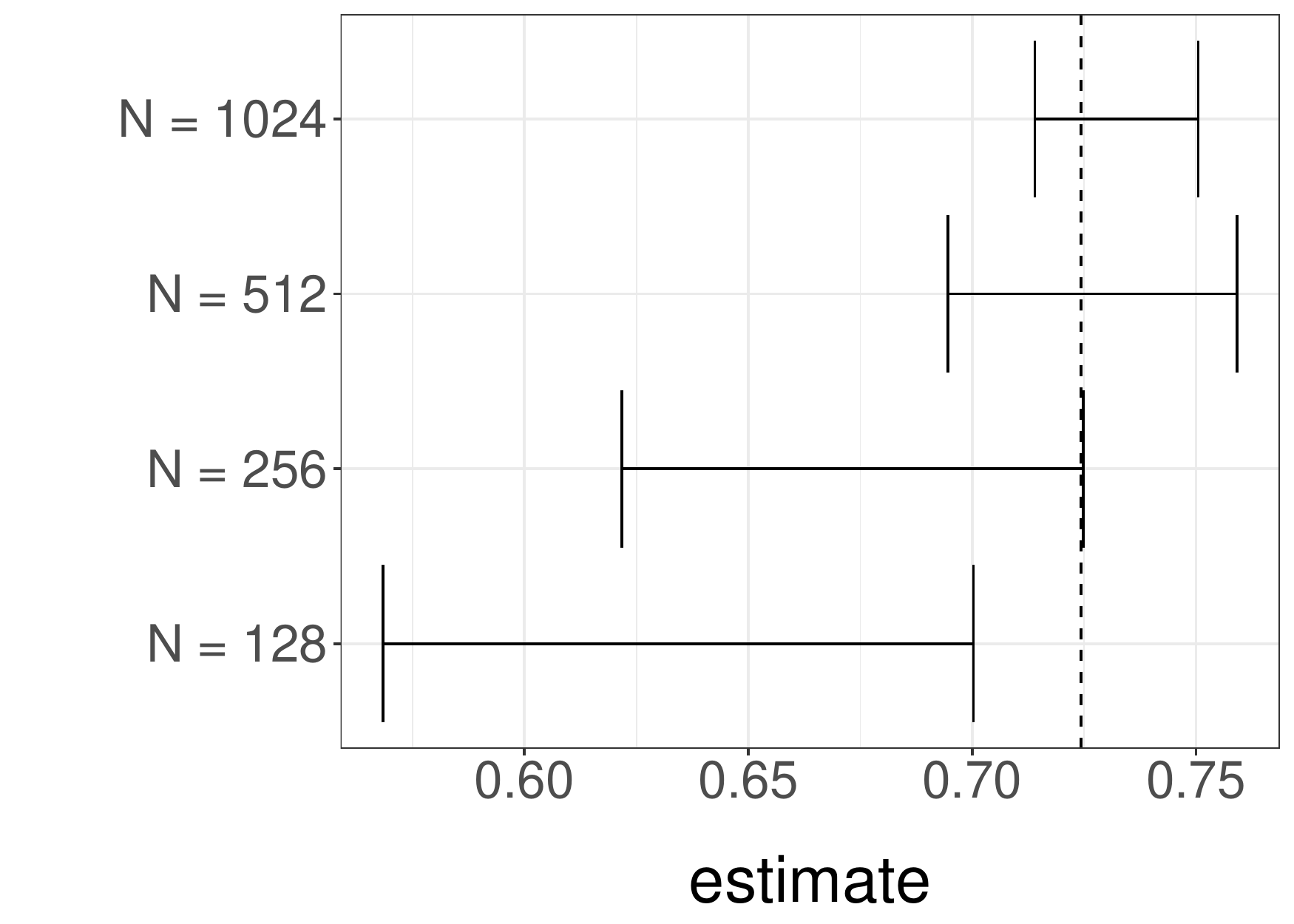}
\par\end{centering}
}
\par\end{centering}
\caption{\label{fig:unlikely} Experiments with the model of Section \ref{sec:numerics:unlikely}.
On the left, kernel density estimates based on $10,000$ runs show the distributions
of estimates of $\mathbb{E}[x_9|y_{10}]$, using particle filters with various $N$.
The exact smoothing mean is represented by a vertical dashed line. 
On the right, $95\%$ confidence intervals constructed from $10,000$ unbiased estimators 
are shown for various $N$.
}
\end{figure}

We consider the task of estimating the smoothing mean $\mathbb{E}[x_9|y_{10}]$.
We run particle filters for different values of $N$, $10,000$ times independently,
and plot kernel density estimators of the distributions of the estimators 
of $\mathbb{E}[x_9|y_{10}]$ in Figure \ref{fig:unlikely:pf}. The dashed vertical line
represents the estimand $\mathbb{E}[x_9|y_{10}]$, obtained analytically. We see that the bias diminishes 
when $N$ increases, but that it is still significant with $N=16,384$ particles.
For any fixed $N$, if we were to ignore the bias and produce confidence intervals 
using the central limit theorem based on independent particle filter estimators, the associated coverage 
would go to zero as the number of independent runs would increase. 

In contrast, confidence intervals obtained with the proposed unbiased estimators
are shown in Figure \ref{fig:unlikely:rg}.
For each value of $N$, the average meeting time was estimated from $100$ independent runs (without ancestor sampling),
and then $k$ was set to that estimate, and $m$ equal to $k$. Then, $R=10,000$ 
independent estimators were produced, and confidence intervals were computed as described in Section \ref{sec:newsmoother:practical}. This leads to 
precise intervals for each choice of $N$. The average costs
associated with $N=128$, $N=256$, $N=512$ and $N=1024$ were respectively
matching the costs of particle filters with $3814$, $4952$, $9152$ and $13,762$ particles.
To conclude, if we match computational costs and compare mean squared errors,
the proposed method is not necessarily advantageous. However, if the interest lies 
in confidence intervals with adequate coverage, the proposed approach comes with 
guarantees thanks to the lack of bias and the central limit theorem
for i.i.d. variables.

\subsection{Prey-predator model \label{sec:numerics:pz}}

Our last example involves  
a model of plankton--zooplankton dynamics taken from \citet{jones2010bayesian}, in which
the transition density is intractable \citep{breto2009time,jacob2015sequential}. 
The bootstrap particle filter is still implementable,
and one can either keep the entire trajectories of the particle filter,
or perform fixed-lag approximations to perform smoothing. On the other hand, backward and ancestor sampling are not implementable.

The hidden state $x_t = (p_t, z_t)$ represents the population size of phytoplankton and zooplankton, and the transition from time $t$ to $t+1$
is given by a Lotka--Volterra equation,
\[\frac{dp_t}{dt} = \alpha p_t - c p_t z_t , \quad \text{and}\quad \frac{dz_t}{dt} = e c p_t z_t -m_l z_t -m_q z_t^2,\]
where the stochastic daily growth rate $\alpha$ is drawn from $\mathcal{N}(\mu_\alpha,\sigma_\alpha^2)$ at every integer time~$t$.
The propagation of each particle involves solving the
above equation numerically using a Runge-Kutta method in  
the \texttt{odeint} library \citep{ahnert2011odeint}. 
The initial distribution is given by
$\log p_0  \sim \mathcal{N}(\log 2 , 1)$ and $\log z_0  \sim \mathcal{N}(\log 2, 1)$.
The parameters $c$ and $e$ represent the clearance rate of the prey and the
growth efficiency of the predator. Both $m_l$ and $m_q$ parameterize the  
mortality rate of the predator.  
The observations $y_t$ are noisy measurements of the phytoplankton $p_t$, $\log y_t \sim \mathcal{N}(\log
p_t, 0.2^2)$; $z_t$ is not observed.
We generate $T = 365$ observations using $\mu_\alpha = 0.7, \sigma_\alpha = 0.5$, 
$c = 0.25$, $e = 0.3$, $m_l = 0.1$, $m_q = 0.1$. 
We consider the problem of estimating the mean population of zooplankton at each time $t\in0:T$,
denoted by $\mathbb{E}[z_t|y_{1:T}]$, given the data-generating parameter.

\begin{figure}
\begin{centering}
\subfloat[\label{fig:pz:meetings} Meeting times.]{\begin{centering}
    \includegraphics[width=0.325\textwidth]{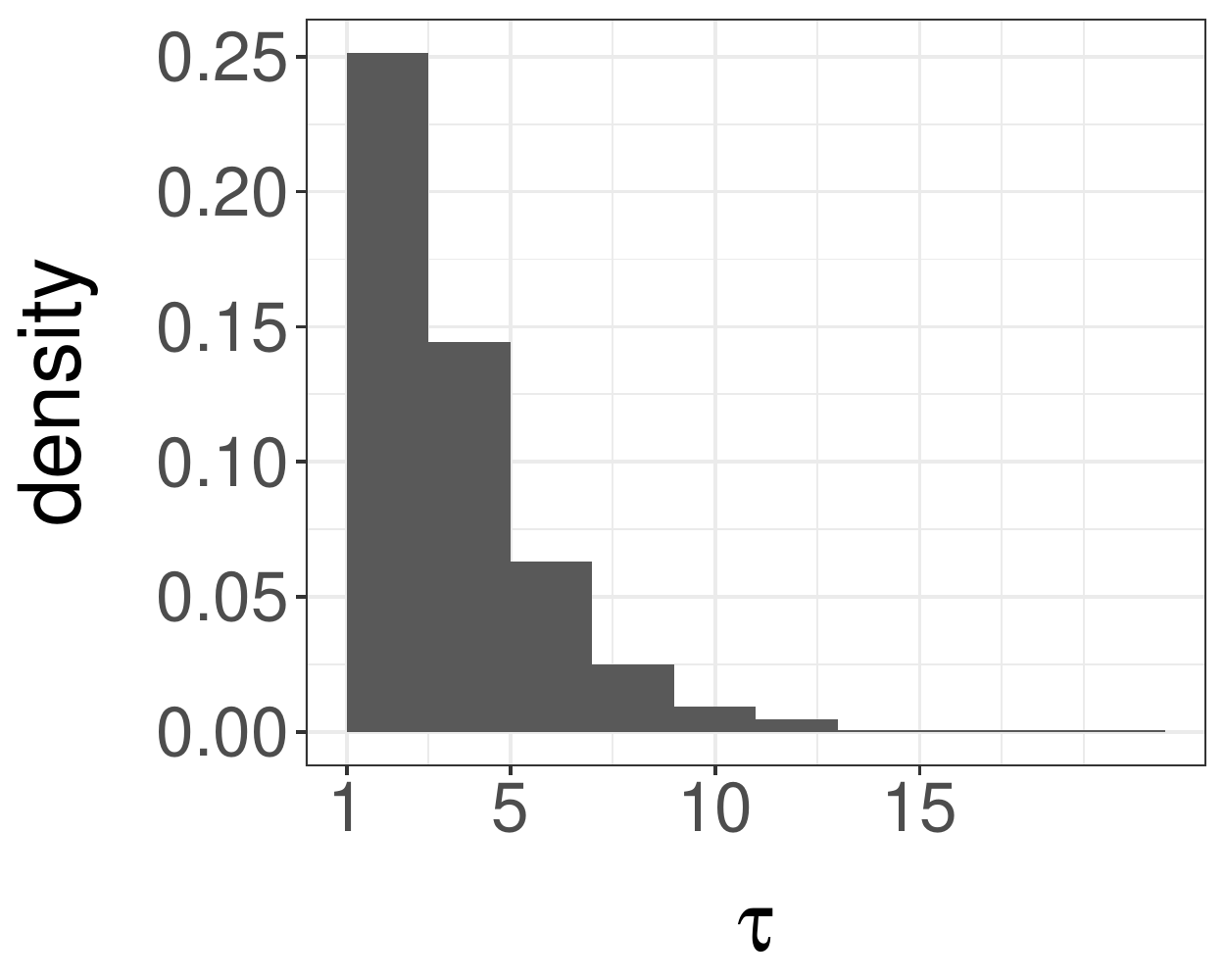}
\par\end{centering}
}
\subfloat[\label{fig:pz:relvar}  Relative variance of smoothing mean estimators,
for the population of zooplankton at each time.]{\begin{centering}
    \includegraphics[width=0.65\textwidth]{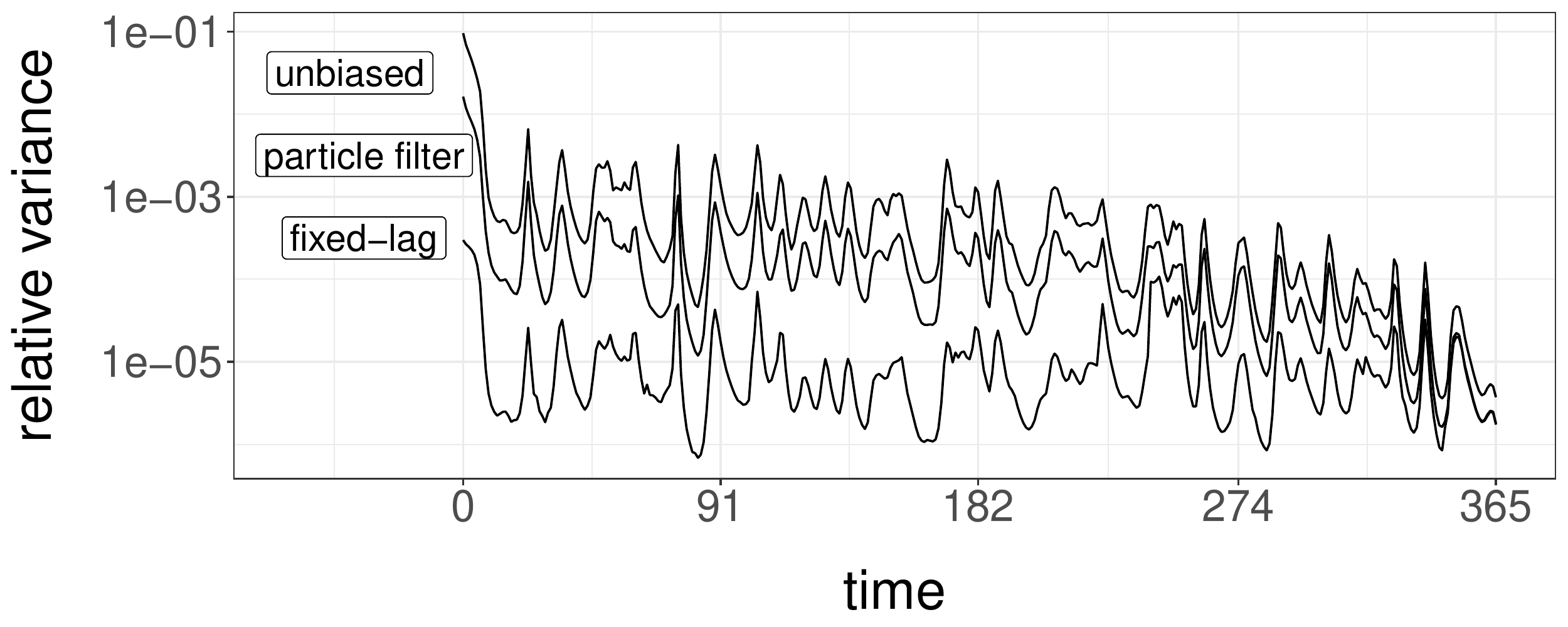}
\par\end{centering}
}
\par\end{centering}
\caption{\label{fig:pz} Experiments with phytoplankton-zooplankton model of Section \ref{sec:numerics:pz}.
On the left, histogram of $1,000$ independent meeting times, obtained with CCPF chains using $N=4,096$ particles.
On the right, relative variance of estimators of $\mathbb{E}[z_t|y_{1:T}]$ for all $t$.
The proposed unbiased estimators (``unbiased'') use $N=4,096$, $k=7$, $m=14$; the particle filters
(``particle filter'') and fixed-lag smoothers (``fixed-lag'') use $N=65,536$, which makes all costs comparable
in terms of numbers of particle propagations and weight evaluations.}
\end{figure}

The distribution of meeting times obtained with $N=4,096$ particles
over $R=1,000$ experiments is shown in
Figure \ref{fig:pz:meetings}. Based on this graph, we choose $k=7$, $m=2k=14$,
and produce $R=1,000$ independent estimators of the smoothing means 
$\mathbb{E}[z_t|y_{1:T}]$. We compute the smoothing means 
with a long CPF chain, taken as ground truth. We then compute the relative variance
of our estimators, defined as their variance divided by the square of the smoothing means.
We find the average cost of the proposed estimator to be equivalent to 
that of a particle filter with $78,377$ particles. To approximately match the cost,
we thus run particle filters with $2^{16}=65,536$ particles, with and without
fixed-lag smoothing with a lag of $10$. The resulting relative variances
are shown in Figure \ref{fig:pz:relvar}. We see that the proposed estimators
yield a larger variance than particle filters, but that the difference is 
manageable. Fixed-lag smoothing provides significant variance reduction,
particularly for earlier time indices.  We can also verify
that the bias of fixed-lag smoothing is negligible in the present example; this would however be hard to assess
with fixed-lag smoothers alone.

\section{Discussion\label{sec:discussion}}

The performance of the proposed estimator is tied to the meeting time. As in \citet{ChopinS:2015}, the
coupling inequality \citep{lindvall2002lectures} can be used to relate  the
meeting time with the mixing of the underlying conditional particle filter
kernel. The proposed approach can be seen as a framework to
parallelize CPF chains and to obtain reliable confidence intervals
over independent replicates.
Any improvement in the CPF directly translates into 
more efficient Rhee--Glynn estimators, as we have illustrated in Section \ref{sec:numerics:hiddenar} with auxiliary particle filters
and ancestor sampling. The methods proposed e.g. in \citet{SinghLM:2017,del2015sequential,guarniero2015iterated,gerber2015sequential,heng2017controlled}
could also be used in Rhee--Glynn estimators, with the hope of obtaining shorter meeting times and smaller variance.

We have considered the estimation of latent processes
given known parameters. In the case of unknown parameters,
joint inference of parameters and latent processes can be done with MCMC methods,
and particle MCMC methods in particular
\citep{andrieu:doucet:holenstein:2010}. Couplings of generic particle MCMC methods
could be achieved by combining couplings proposed in the 
present article with those described in \citet{jacob2017unbiased} for Metropolis--Hastings chains. 
Furthermore, for fixed parameters, coupling the particle independent Metropolis--Hastings algorithm 
of \citet{andrieu:doucet:holenstein:2010} would lead to unbiased estimators
of smoothing expectations that would not require coupled resampling schemes (see Section \ref{sec:couplingparticlesystems}). 

The appeal of the proposed smoother, namely parallelization over independent replicates and 
confidence intervals, would be shared by perfect samplers. These algorithms aim at the more 
ambitious task of sampling exactly from the smoothing distribution \citep{leedoucetperfectsimulation}.
It remains unknown whether the proposed approach could play a role in the design of perfect samplers.
We have established the validity of the Rhee--Glynn estimator under mild conditions, 
but its theoretical study as a function of the
time horizon and the number of particles deserves further analysis \citep[see][for a path forward]{Lee2018ccbpf}. 
Finally, together with Fisher's
identity \citep{douc:moulines:2014}, the proposed smoother provides unbiased estimators of the
score for models where the transition density is tractable. This could help maximizing the 
likelihood via stochastic gradient ascent.

    \textbf{Acknowledgements.}
The authors thank Marco Cuturi, Mathieu Gerber, Jeremy Heng and Anthony Lee for
helpful discussions. This work was initiated during the workshop on
\emph{Advanced Monte Carlo methods for complex inference problems} at
the Isaac Newton Institute for Mathematical Sciences, Cambridge, UK
held in April 2014. We would like to thank the organizers for a great
event which led to this work.

\bibliographystyle{apalike}
\bibliography{Biblio}

\appendix

\section{Intermediate result on the meeting probability \label{sec:proof:intermed}}

Before proving Theorem \ref{thm:finitevariance}, we introduce
an intermediate result on the probability of the chains meeting at the next step,
irrespective of their current states.  
The result provides a lower-bound on the probability of meeting in one step,
for coupled chains generated by the coupled conditional particle filter (CCPF) kernel. 

\begin{lemma}
    Let $N\geq 2$ and $T\geq 1$ be fixed.  Under Assumptions
    \ref{assumption:upperbound} and \ref{assumption:couplingmatrix},  there
    exists $\varepsilon>0$, depending on $N$ and $T$, such that 
\[
    \forall X  \in \mathbb{X}^{T+1}, \quad \forall \tX \in \mathbb{X}^{T+1}, \quad \mathbb{P}(X' = \tX' | X, \tX) \geq \varepsilon,
\]
where $(X',\tX') \sim \text{CCPF}((X,\tX), \cdot)$.
Furthermore, if $X = \tX$, then $X' = \tX'$ almost surely.
    \label{lemma:meetingprobability}
\end{lemma}
The constant $\varepsilon$
depends on $N$ and $T$, and on the coupled resampling scheme being used.  Lemma
\ref{lemma:meetingprobability} can be used, together with the coupling
inequality \citep{lindvall2002lectures}, to prove the ergodicity of the
conditional particle filter kernel, which is akin to the approach of
\citet{ChopinS:2015}.  The coupling inequality states that the total variation
distance between $X^{(n)}$ and  $\tX^{(n-1)}$ is less than
$2\mathbb{P}(\tau > n)$, where $\tau$ is the meeting time.  By assuming $\tX^{(0)}\sim\pi$,
$\tX^{(n)}$ follows $\pi$ at each step $n$, and we obtain a bound for the total
variation distance between $X^{(n)}$ and $\pi$.  Using Lemma
\ref{lemma:meetingprobability}, we can bound the probability
$\mathbb{P}(\tau > n)$ from above by $(1-\varepsilon)^n$, as in the proof of Theorem \ref{thm:finitevariance} below.
This implies that the computational cost of the proposed estimator
has a finite expectation for all $N\geq 2$ and~$T\geq 1$.  

\emph{Proof of Lemma \ref{lemma:meetingprobability}}.
We write $\PrbX{t}$ and $\EX{t}$ for the conditional probability and expectation, respectively, 
with respect to the law of the particles generated by the CCPF procedure conditionally on the reference trajectories up to time $t$, $(x_{0:t}, \tilde x_{0:t})$.
Furthermore, let $\mathcal{F}_t$ denote the filtrations generated by the CCPF at time $t$.  We denote by $x_{0:t}^k$, for
$k\in1:N$, the surviving trajectories at time~$t$.
Let $I_t \subseteq 1:N-1$ be the set of common particles at time $t$ defined by
    $I_t = \{j \in 1:N-1 : x_{0:t}^j  = \tilde x_{0:t}^j \}$.
	The meeting probability can then be bounded by:
\begin{multline}
    \PrbX{T}(x_{0:T}^\prime = \tilde x_{0:T}^\prime) = \EX{T}\left[\I\!\left(x_{0:T}^{b_T} = \tilde x_{0:T}^{\tilde{b}_T} \right)\right]
	\geq \sum_{k=1}^{N-1} \EX{T}[\I\!\left(k \in I_T\right) P_T^{kk}] \\
	= (N-1)\EX{T}[\I\!\left(1\in I_T \right) P_T^{11}]
	\geq \frac{N-1}{ (N\bar{g})^2} \EX{T}[\I\!\left(1\in I_T \right) g_T(x_T^1) g_T(\tilde x_T^1)],
\end{multline}
where we have used Assumptions \ref{assumption:upperbound} and \ref{assumption:couplingmatrix}.

Now, let $\psi_t : \setX^t \mapsto \reals_+$ and consider
\begin{align}
	\label{eq:crude:h}
	\EX{t}[\I\!\left( 1\in I_t \right) \psi_t(x_{0:t}^1) \psi_t(\tilde x_{0:t}^1)] =
	\EX{t}[\I\!\left( 1\in I_t \right) \psi_t(x_{0:t}^1)^2],
\end{align}
since the two trajectories agree on $\{1\in I_t\}$.
We have
\begin{align}
	\I\!\left( 1\in I_t \right) \geq \sum_{k=1}^{N-1} \I\!\left(k\in I_{t-1} \right) \I\!\left(a_{t-1}^1 = \tilde a_{t-1}^1 = k \right),
\end{align}
and thus
\begin{multline}
	\label{eq:crude:h2}
	\EX{t}[\I\!\left( 1\in I_t \right) \psi_t(x_{0:t}^1)^2] \\
	\geq \EX{t}[\sum_{k=1}^{N-1} \I\!\left(k\in I_{t-1} \right) \EX{t}[ \I\!\left(a_{t-1}^1 = \tilde a_{t-1}^1 = k \right) \psi_t(x_{0:t}^1)^2 \mid \mathcal{F}_{t-1} ]] \\
	= (N-1)\EX{t}[\I\!\left(1\in I_{t-1} \right) \EX{t}[ \I\!\left(a_{t-1}^1 = \tilde a_{t-1}^1 = 1 \right) \psi_t(x_{0:t}^1)^2 \mid \mathcal{F}_{t-1}  ]].
\end{multline}
The inner conditional expectation can be computed as
\begin{multline}
	\label{eq:cruce:h2-inner}
	\EX{t}[ \I\!\left(a_{t-1}^1 = \tilde a_{t-1}^1 = 1 \right) \psi_t(x_{0:t}^1)^2 \mid \mathcal{F}_{t-1} ] \\
	=\sum_{k,\ell=1}^N P_{t-1}^{k\ell} \I\!\left(k=\ell=1\right) \int \psi_t((x_{0:t-1}^k, x_t ))^2 f(dx_t|x_{t-1}^k) \\
	= P_{t-1}^{11} \int \psi_t((x_{0:t-1}^1, x_t))^2 f(dx_t|x_{t-1}^1) \\
    \geq \frac{g_{t-1}(x_{t-1}^1) g_{t-1}(\tilde x_{t-1}^1) }{(N\bar{g})^2} \left( \int \psi_t((x_{0:t-1}^1, x_t )) f(dx_t|x_{t-1}^1) \right)^2,
\end{multline}
where we have again used Assumptions \ref{assumption:upperbound} and \ref{assumption:couplingmatrix}.
Note that this expression is independent of the final states of the reference trajectories, $(x_t, \tilde x_t)$, which can thus be dropped from the conditioning.
Furthermore, on $\{1\in I_{t-1}\}$ it holds that $x_{0:t-1}^1 = \tilde x_{0:t-1}^1$ and therefore, combining Eqs.~\eqref{eq:crude:h}--\eqref{eq:cruce:h2-inner} we get
\begin{multline}
	\EX{t}[\I\!\left( 1\in I_t \right) \psi_t(x_{0:t}^1) \psi_t(\tilde x_{0:t}^1)] \\
	\geq \frac{(N-1)}{(N\bar{g})^2}\EX{t-1}\Big[\I\!\left(1\in I_{t-1} \right) g_{t-1}(x_{t-1}^1) \int \psi_t((x_{0:t-1}^1, x_t )) f(dx_t|x_{t-1}^1) \\ \times
	g_{t-1}(\tilde x_{t-1}^1) \int \psi_t((\tilde x_{0:t-1}^1, x_t )) f(dx_t|\tilde x_{t-1}^1)
	\Big].
\end{multline}
Thus, if we define
for $t=1,\ldots,T-1$, 
$\psi_t(x_{0:t}) = g_t(x_t) \int \psi_{t+1}(x_{0:t+1}) f(dx_{t+1}|x_t)$,  
and
$\psi_T(x_{0:T}) = g_T(x_T)$, 
it follows that
\begin{align*}
	\PrbX{T}(x_{0:T}^\prime= \tilde x_{0:T}^\prime) &\geq \frac{(N-1)^\transp}{(N\bar{g})^{2T}} \EXzero[\I\!\left(1\in I_1 \right) \psi_1(x_1^1)\psi_1(\tilde x_1^1)] \\
	&= \frac{(N-1)^\transp}{(N\bar{g})^{2T}} \EXzero[\psi_1(x_1^1)^2] \geq \frac{(N-1)^\transp}{(N\bar{g})^{2T}} Z^2 > 0,
\end{align*}
where $Z > 0$ is the normalizing constant of the model, 
$Z=\int m_0(dx_0) \prod_{t=1}^\transp g_t(x_t) f(dx_t|x_{t-1})$. 
This concludes the proof of Lemma \ref{lemma:meetingprobability}.

For any fixed $T$, the bound goes to zero when $N\to \infty$.
The proof fails to capture accurately the behaviour of $\varepsilon$
in Lemma \ref{lemma:meetingprobability} as a function of $N$ and $T$.
Indeed, we observe in the numerical experiments of Section \ref{sec:numerics} that meeting times decrease 
when $N$ increases.

\section{Proof of Theorem \ref{thm:finitevariance} \label{sec:proof:unbiased}}

The proof is similar to those presented in \citet{rhee:phd}, in \citet{McLeish:2011},
\citet{vihola2015unbiased}, and \citet{glynn2014exact}.
We can first upper-bound $\mathbb{P}\left(\tau>n\right)$, for all $n\geq2$,
using Lemma \ref{lemma:meetingprobability} \citep[e.g.][exercise E.10.5]{williams1991probability}.
We obtain for all $n\geq2$,
\begin{equation}
\mathbb{P}\left(\tau>n\right)\leq\left(1-\varepsilon\right)^{n-1}.\label{eq:meetingtime:survival2}
\end{equation}
This ensures that $\mathbb{E}[\tau]$ is finite; and that $\tau$ is almost surely finite.
We then introduce the random variables $Z_{m}=\sum_{n=0}^{m} \Delta^{(n)}$ for all $m\geq 1$.
Since $\tau$ is almost surely finite, and since $\Delta^{(n)} = 0$ for all $n \geq \tau$,
then $Z_m\to Z_\tau = H_0$ almost surely when $m\to\infty$. 
We prove that $(Z_m)_{m\geq 1}$ is a Cauchy sequence in 
$L_2$, i.e.
    $\sup_{m'\geq m} \mathbb{E}\left[ (Z_{m'} - Z_m)^2 \right]$
    goes to $0$ as $m\to\infty$. 
We write 
\begin{align}
    \label{eq:zcauchy}
    \mathbb{E}[(Z_{m'} - Z_m)^2] &= \sum_{n = m + 1}^{m'}\sum_{\ell = m + 1}^{m'} \mathbb{E}[\Delta^{(n)}\Delta^{(\ell)}].
\end{align}
We use Cauchy-Schwarz inequality to write
$(\mathbb{E}[\Delta^{(n)}\Delta^{(\ell)}])^2 \leq \mathbb{E}[(\Delta^{(n)})^2]\mathbb{E}[(\Delta^{(\ell)})^2]$,
and we note that $(\Delta^{(n)})^2= \Delta^{(n)}\mathds{1}(\tau>n)$. Together 
with H\"older's inequality with $p=1+\delta/2$, and $q=(2+\delta)/\delta$,
where $\delta$ is as in Assumption \ref{assumption:mixing}, 
we can write 
\begin{align*}
\mathbb{E}\left[(\Delta^{(n)})^{2}\right]  & \leq\mathbb{E}\left[(\Delta^{(n)})^{2+\delta}\right]^{1/(1+\delta/2)}\left(\left(1-\varepsilon\right)^{\delta/(2+\delta)}\right)^{n-1}.
\end{align*}
Furthermore, using Assumption \ref{assumption:mixing}
and Minkowski's inequality, we obtain the bound 
\begin{align*}
     \forall n\geq n_0, \qquad & \mathbb{E}\left[(\Delta^{(n)})^{2+\delta}\right]^{1/(1+\delta/2)}\leq C_{1},
\end{align*}
where $C_1$ is independent of $n$. The above inequalities lead to 
the terms $\mathbb{E}[\Delta^{(n)}\Delta^{(\ell)}]$ being upper bounded by an expression of the form $C_1 \eta^n \eta^\ell$,
where $\eta \in (0,1)$. 
Thus we can compute a bound on Eq. \eqref{eq:zcauchy}, by computing geometric series, 
and finally conclude that $(Z_m)_{m \geq 1}$ is a Cauchy sequence in $L_2$.

By uniqueness of the limit, since $(Z_m)_{m \geq 1}$ goes almost surely to
$H_0$, $(Z_m)_{m \geq 1}$ goes to $H_0$ in $L_2$. This shows that $H_0$ has finite first two moments. We can retrieve the
expectation of $H_0$ by
\[ \mathbb{E}Z_{m}=\sum_{n=0}^{m}\mathbb{E}[\Delta^{(n)}]=\mathbb{E}\left[h(X^{(m)})\right] \xrightarrow[m\to \infty]{} \pi(h),
\]
according to Assumption \ref{assumption:mixing}. This concludes the proof of Theorem \ref{thm:finitevariance}
for $H_k$ with $k=0$, and a similar reasoning applies for any $k\geq 0$.

\end{document}

%% file: tablear1horizon.tex
\begin{table}[ht]
\centering
\begin{tabular}{llllll}
  &  & \multicolumn{2}{c}{Bootstrap PF}  & \multicolumn{2}{c}{Auxiliary PF}  \\ 
 & & without AS & with AS & without AS & with AS \\ 
  \hline
  N = 128& T = 50 & 17.84 (17.13) & 7.73 (5.11) & 3.96 (2.3) & 3.37 (1.42) \\ 
  N = 256& T = 100 & 13.16 (11.09) & 7.59 (5.05) & 3.78 (1.99) & 3.16 (1.09) \\ 
  N = 512& T = 200 & 12.52 (10.64) & 6.77 (3.85) & 3.52 (1.75) & 2.97 (0.94) \\ 
  N = 1024& T = 400 & 12.74 (10.96) & 6.77 (3.47) & 3.69 (1.94) & 2.91 (0.87) \\ 
  N = 2048& T = 800 & 13.58 (9.56) & 6.34 (2.95) & 3.54 (1.9) & 2.95 (0.87) \\ 
\end{tabular}
\caption{Average meeting time, as a function of the number of particles $N$ and the time
horizon $T$, with bootstrap particle filters and auxiliary particle filters, 
with and without ancestor sampling (AS), computed over $R=500$ experiments.
Standard deviations are between brackets. Results obtained in the hidden auto-regressive model
of Section \ref{sec:numerics:hiddenar}.
\label{table:effecthorizon}} 
\end{table}